\documentclass[12pt,a4paper]{article}
\usepackage{graphicx}

\def\qed{{\unskip\nobreak\hfil\penalty50\hskip .001pt \hbox{}\nobreak\hfil
          \vrule height 1.2ex width 1.1ex depth -.1ex
          \parfillskip=0pt\finalhyphendemerits=0\medbreak}\rm}

\def\Theorem #1. {\bigbreak\vskip-\parskip\noindent{\bf  Theorem   #1.}
    \quad\it}
\def\Corollary #1. {\bigbreak\vskip-\parskip\noindent{\bf Corollary #1.}
   \quad\it}

\def\Proof#1.{\rm\par\ifdim\lastskip<\bigskipamount\removelastskip\fi
    \smallskip\noindent{\bf Proof.}\quad}
\def\R{{{\rm I} \! {\rm R}}}
\def\C{{{\rm I} \! \hspace*{-1.3mm} {\rm C}}}

\begin{document}
\vspace*{2cm}
\begin{center}
 \Huge\bf
Ghost spinors in quantum particles interference
\vspace*{0.25in}

\large

Elena V. Palesheva
\vspace*{0.15in}

\normalsize

Department of Mathematics, Omsk State University \\
644077 Omsk-77 RUSSIA
\\
\vspace*{0.5cm}
E-mail: m82palesheva@math.omsu.omskreg.ru  \\
\vspace*{0.5cm}
\vspace{.5in}
ABSTRACT
\end{center}
In this article a question of the ghost spinors influence to the
quantum particles interference is investigated. The interaction between
spinors and ghost spinors are considered. Furthermore the conditions of
zero-point
energy-momentum tensor in private cases are found. Also we consider a question
about the experimental test of Deutsch shadow pacticles existense.

\newpage

\setcounter{page}{1}


\section*{Introduction}
In this article we consider some development of Deutsch idea \cite{2}.
David Deutsch consider the known experiment of quantum mechanics, namely the
experiment with a quantum particles interference, and then
the received interference pattern is explained by shadow particles
existence, there are particles in parallel universes. The
offered approach to description of our
reality \cite{2} get a mathematical motivation in \cite{3,4}.
The Guts-Deutsch Multiverse contain set of parallel universes, herewith, as David
Deutsch expect \cite{2}, particles in our universe can interact only with own
shadow particles. The sintetical differential geometry is used in Guts model
of Multiverse. As ghost spinors have a zero-point energy-momentum tensor,
identification of Deutsch shadow particles in case of spinor fields and ghost spinors
was made in \cite{11,12}. In this article we consider a question of
ghost spinors influence to interference pattern. Herewith in first we
consider some conditions of zero-point energy-momentum tensor which will be used
hereinafter.

\section{Zero-point energy-momentum tensor}

We consider only a spacetime of special relativity.
As known the Dirac equation for free particle in spacetime of Minkowski
hase the next form
\begin{equation}\label{d}
i\hbar {\gamma}^{(k)}\frac{\partial\psi}{\partial x^{\scriptscriptstyle
k}}-mc\psi =0\, ,
\end{equation}
where ${\gamma}^{(k)}$ are Dirac matrixes in standart presentation:
$$
{\gamma}^{(0)}=\left[\begin{array}{cc}I&0\\
0&-I\end{array}\right]\, ,\quad{\gamma}^{(\alpha)}=
\left[\begin{array}{cc}0&{\sigma}_{\alpha}\\
-{\sigma}_{\alpha}&0\end{array}\right]\, ,
$$
$$
{\sigma}_1=\left[\begin{array}{cc}0&1\\ 1&0\end{array}\right]\, ,
{\sigma}_2=\left[\begin{array}{cc}0&-i\\ i&0\end{array}\right]\, ,
{\sigma}_3=\left[\begin{array}{cc}1&0\\ 0&-1\end{array}\right]\, ,
I=\left[\begin{array}{cc}1&0\\ 0&1\end{array}\right]\, .
$$
In this case the energy-momentum tensor of spinor field is defined by
expression
\begin{equation}\label{t}
T_{ik}=\frac{i\hbar
c}{4}\left\{{\psi}^*{\gamma}^{(0)}{\gamma}_i\frac{\partial\psi}{\partial
x^{\scriptscriptstyle k}}-\frac{\partial {\psi}^*}{\partial
x^{\scriptscriptstyle k}}{\gamma}^{(0)}{\gamma}_i\psi+{\psi}
^*{\gamma}^{(0)}{\gamma}_k\frac{\partial\psi}{\partial x^{\scriptscriptstyle
i}} -\frac{\partial {\psi}^*}{\partial x^{\scriptscriptstyle
i}}{\gamma}^{(0)} {\gamma}_k\psi\right\}\, .
\end{equation}
Herewith the symbol ${}^*$ means Hermite conjugation and
$$
{\gamma}_i=g_{ik}{\gamma}^{(k)}.
$$
In \cite{6,7,8,9,11,12} the solutions of Dirac equation with zero-point
energy-momentum tensor and non-zero-point current density were found. The
formula
\begin{equation}\label{8}
j^{(k)}={\psi}^*{\gamma}^{(0)}{\gamma}^{(k)}\psi
\end{equation}
define current density in Minkowski spacetime.
Herewith $j^{(0)}$ is a square of modulus of probability amplitude
$\psi$
which characterize the probability of appearance of this particle in spacetime.
In special relativity $j^{(0)}={\psi}^*\psi$ and $j^{(\alpha)}$ is
a  velosity of change of probability density
\footnote{Greek indexes are 1,2,3.}.

\Theorem 1.
Let $\psi=u\cdot G(x)$ is a solution of Dirac equation herewith
$$
{\psi}^*\psi\neq 0,\quad G(x)=f(x) +i\cdot g(x),
$$
where $f(x)$ and $g(x)$ are real smooth functions and bispinor
$$
u=\left[\begin{array}{l} u_0\\ u_1\\ u_2\\
u_3\end{array}\right]
$$
such that $\forall\; i\quad u_i\in \C.$ In these conditions
$\psi$ is a ghost spinor iff $g(x)=a\cdot f(x)$ where $a=const\in \R.$

\Proof.
From definition the solution of Dirac equation is a ghost spinor iff
$T_{ik}\equiv 0$ and Dirac current $j^{(k)}\neq 0.$
That  $j^{(k)}\neq 0$ follow from ${\psi}^*\psi\neq 0.$

Now we must show that $T_{ik}\equiv 0$ iff $g(x)=a\cdot f(x).$
For this we notice that~\footnote {Here $\overline{G}$ is a
complex conjugate function.}
$$
{\psi}^*={u}^*\overline{G}={u}^*(f(x)-i\cdot g(x)).
$$
Hereinafter we consider some stages.

a) We notice that identity $T_{oo}\equiv 0$ exist iff
$$
0={\psi}^*{\gamma}^{(0)}{\gamma}^{(0)}\frac{\partial \psi}{\partial
x^{\scriptscriptstyle 0}}-\frac{\partial{\psi}^*}{\partial
x^{\scriptscriptstyle
0}}{\gamma}^{(0)}{\gamma}^{(0)}\psi={\psi}^*\frac{\partial\psi}{\partial
x^{\scriptscriptstyle 0} }-\frac{\partial{\psi}^*}{\partial
x^{\scriptscriptstyle 0}}\psi=u^*\overline{G}u \frac{\partial G}{\partial
x^{\scriptscriptstyle 0}}-
$$
$$-u^*\frac{\partial \overline{G}}{\partial x^
{\scriptscriptstyle 0}}uG=u^*u\left(\overline{G}\frac{\partial G}{\partial
x^{\scriptscriptstyle 0}} -\frac{\partial \overline{G}}{\partial
x^{\scriptscriptstyle 0}}G\right).
$$
As ${\psi}^*\psi\neq 0$ we have $u^*u\neq 0.$ So we get that
$T_{oo}\equiv 0$ iff
$$
\overline{G}\frac{\partial G}{\partial x^
{\scriptscriptstyle 0}}-\frac{\partial \overline{G}}{\partial
x^{\scriptscriptstyle 0}}G=0\, .
$$
That is right iff
$$
f\frac{\partial g}{\partial x^{ \scriptscriptstyle 0}}=g\frac{\partial
f}{\partial x^{\scriptscriptstyle 0}}.
$$

b) Now we consider a condition on $T_{01}.$ We have that
$T_{01}\equiv 0$ iff the next equality is right:
$$
0={\psi}^*{\gamma}^{(0)}{\gamma}^{(0)}\frac{\partial \psi}{\partial
x^{\scriptscriptstyle 1}}-\frac{\partial{\psi}^*}{\partial
x^{\scriptscriptstyle
1}}{\gamma}^{(0)}{\gamma}^{(0)}\psi-{\psi}^*{\gamma}^{(0)}{\gamma}^{(1)}\frac{\partial
\psi}{\partial x^{\scriptscriptstyle 1}}+\frac{\partial{\psi}^*}{\partial
x^{\scriptscriptstyle
1}}{\gamma}^{(0)}{\gamma}^{(1)}\psi=
$$
$$
={\psi}^*\frac{\partial\psi}{\partial
x^{\scriptscriptstyle 1}}-\frac{\partial{\psi}^*}{\partial
x^{\scriptscriptstyle 1}}\psi-u^*\overline{G}{\gamma}^
{(0)}{\gamma}^{(1)}u\frac{\partial G}{\partial x^{\scriptscriptstyle
0}}+u^*\frac{\partial \overline{G}}{\partial x^{\scriptscriptstyle
0}}{\gamma}^{(0)}{\gamma}^{(1)}uG =
$$
$$
=u^*u\left(\overline{G}\frac{\partial
G}{\partial x^{\scriptscriptstyle 1}} -\frac{\partial \overline{G}}{\partial
x^{\scriptscriptstyle 1}}G\right)-u^*{\gamma}^{(0)}{\gamma}^
{(1)}u\left(\overline{G}\frac{\partial G}{\partial x^{\scriptscriptstyle 0}}
-\frac{\partial \overline{G}}{\partial x^{\scriptscriptstyle 0}}G\right).
$$
From results of stage a) we have that second summand is equal to zero. So $T_{01}$ is
equal to zero iff
$$
f\frac{\partial g}{\partial x^{ \scriptscriptstyle
1}}=g\frac{\partial f}{\partial x^{\scriptscriptstyle 1}}.
$$

c) Let the similar procedure for components $T_{02}$ and
$T_{03}$ is consecutively executed. Then we shall find that $T_{0k}\equiv 0$ iff
$$
f\frac{\partial g}{\partial x^{ \scriptscriptstyle
k}}=g\frac{\partial f}{\partial x^{\scriptscriptstyle k}}.
$$
This sistem hase solution
\begin{equation}\label{af}
g(x)=af(x)\, ,
\end{equation}
where $a=const\in \R.$

Now we shall show that an expression
$$
\forall\; \alpha, \beta\quad T_{\alpha\beta}\equiv 0
$$
follow from (\ref{af}).
We have that $T_{\alpha\beta}\equiv 0$ iff
$$
0={\psi}^*{\gamma}^{(0)}{\gamma}^{(\alpha)}\frac{\partial \psi}{\partial
x^{\scriptscriptstyle \beta}}-\frac{\partial{\psi}^*}{\partial x^
{\scriptscriptstyle
\beta}}{\gamma}^{(0)}{\gamma}^{(\alpha)}\psi+{\psi}^*{\gamma}^{(0)}{\gamma}^{(\beta)}\frac
{\partial\psi}{\partial x^{\scriptscriptstyle
\alpha}}-\frac{\partial{\psi}^*}{\partial x^
{\scriptscriptstyle\alpha}}{\gamma}^{(0)}{\gamma}^{(\beta)}\psi=
$$
$$
=u^*{\gamma}^{(0)}{\gamma}^
{(\alpha)}u\left(\overline{G}\frac{\partial G}{\partial x^{\scriptscriptstyle
\beta}} -\frac{\partial \overline{G}}{\partial x^{\scriptscriptstyle
\beta}}G\right)+u^*{\gamma}^{(0)} {\gamma}^ {(\beta)}u\left(\overline{G}\frac{\partial
G}{\partial x^{\scriptscriptstyle \alpha}} -\frac{\partial
\overline{G}}{\partial x^{\scriptscriptstyle \alpha}}G\right).
$$
For type of function $G(x)$ we have a right equation. Theorem is  proved.
\qed

\Corollary 1.
Let the conditions of theorem 1 is executed and
$$
G(x)=e^{\alpha (x)+i\beta(x)}
$$
where $\alpha(x)$ and $\beta(x)$ are smooth real functions.
Then $\psi$ is a ghost spinor iff $\beta(x)=const\in \R$.
\Proof.
$$
G(x)=e^{\alpha (x)+i\beta(x)}=e^{\alpha (x)}\cos[\beta(x)]+i
e^{\alpha (x)}\sin[\beta(x)].
$$
From theorem 1 we have that $\psi$ is a ghost spinor iff ${\rm ctg}[\beta(x)]=const.$
\qed

\Theorem 2.
Let us assume that
$$
\psi=\left[\begin{array}{l} G_0(x)\\ G_1(x)\\ G_2(x)\\
G_3(x)\end{array}\right]
$$
is a solution of Dirac equation, herewith
${\psi}^*\psi\neq 0$ and
$$
\forall\; k \quad G_k(x)=f_k(x) +ig_k(x)
$$
where $f_k(x)$ and $g_k(x)$ are smooth real functions. If
$$
\forall\; i,k\quad
f_i(x)=c_{ik}\cdot g_k(x),
$$
here $c_{ik}=const\in \R$, then
$\psi$ is a ghost spinor.

\Proof.
For considering spinor the next equality is corrected
$$
{\psi}^*{\gamma}^{(0)}{\gamma}_{i}\frac{\partial \psi}{\partial
x^{\scriptscriptstyle k}}-\frac{\partial{\psi}^*}{\partial x^
{\scriptscriptstyle k}}{\gamma}^{(0)}{\gamma}_{i}\psi=0.
$$
So $T_{ik}\equiv 0$. As ${\psi}^*\psi\neq 0$ the Dirac current is not equal
to zero. Theorem is  proved. \qed

\Corollary 2.
Let we have a solution of Dirac equation
\begin{equation}\label{sl2}
\psi=uf(x),
\end{equation}
where $f(x)$ is smooth real function and components of bispinor $u$
are complex numbers such that a statement ${\psi}^*\psi\neq 0$ is right.
Then $\psi$ is a ghost spinor.

\Proof.
Proof is  obviously.
\qed

\section{Interactions between real spinors and ghost spinors}

David Deutsch assume that shadow electrons act to appearance
of interference pattern \cite{2}. A question about nature of interactions
between real and shadow particles is appeared. Deutsch  expect that shadow
photons interact only with own real photons. Here we will describe that if a real spinor wave
and a ghost spinor wave are interacted we observe the interference pattern. Moreover
in some cases of these interactions the resulting wave is a ghost spinor.
Also we will show that if we take into account shadow spinor waves in
the two-slit experiment then it is not disagree to known experimental data.

\subsection{Fluctuation of matter in universe}
Let the Dirac wave function hase a type
$$
\psi =u(f(x)+ig(x)),
$$
$ f(x)$ and $g(x)$ are smooth real functions with conditions
$$
f(x)\neq const\cdot g(x),\quad {[g(x)+f(x)]}^2\neq 0.
$$
Herewith we consider a spinor
$$
u=\left[\begin{array}
{l}u_0\\ u_1\\ u_2\\ u_3\end{array}\right]
$$
such that $\forall\; i\quad u_i \in \R$ and $u^*u\neq 0.$
From theorem~1 we have that such wave is not a ghost spinor.

Let us consider a new particle in state $\theta=u[g(x)-f(x)]$ which also is
a solution of Dirac equation. From results of theorem~1 we notice that
this spinor is not a ghost spinor. Let us assume that $\psi$ and $\theta$
such that in some point of spacetime these waves interact. Then after this
interaction the resulting wave is definded by state $\psi
+\theta=u(1+i)g(x)$. Once agane we use theorem~1 and receive that $\psi
+\theta$ is a ghost spinor.

In result we received that sometimes a shadow particle can interact with
a real particle and after this a shadow wave is received. In general case we
have that sometimes a real electron move over to a shadow state without collision
with another real particle. But a shadow electron is an electron in a parallel
universe \cite{2}. From these results we have that particles in our universe
can "spontaneously disappear" and "spontaneously apper". Herewith
fluctuations of matter is not arise from interaction of particles in one
universe, i.e. fluctuations of matter may be  caused namelly by
interaction of different universes of a Multiverse which model was
presented in \cite{3,4}.

\subsection{Interference between real spinors and ghost spinors}

Let us assume that the solution of Dirac equation (\ref{d}) is
\begin{equation}\label{0}
\psi=\left[\begin{array}{l} u_0\\ u_1\\ u_2\\ u_3\end{array}\right].
\end{equation}
Then insert (\ref{0}) into (\ref{d}) we get the next system of the linear
differential equations of first order in partial derivatives:
$$
\left\{\begin{array}{l} {\displaystyle
u_{0,0}+u_{3,1}-iu_{3,2}+u_{2,3}=-i\frac{mc}{\mathstrut\hbar}u_0}\, ,\\
{\displaystyle u_{1,0}+u_{2,1}+iu_{2,2}-u_{3,3}=-i\frac{\mathstrut
mc}{\mathstrut\hbar}u_1}\, ,\\
{\displaystyle
-u_{2,0}-u_{1,1}+iu_{1,2}-u_{0,3}=-i\frac{mc\mathstrut}{\mathstrut\hbar}u_2}\, ,\\
{\displaystyle
-u_{3,0}-u_{0,1}-iu_{0,2}+u_{1,3}=-i\frac{mc\mathstrut}{\hbar}u_3}\, .
\end{array}\right.
$$
Here $u_{i,k}$ is a differentiation from k-coordinate. Let our solution
satisfy to $u_0=u_1=-u_2=u_3=u$. More over we assume that
$$
\frac{\partial u}{\partial x^{\scriptscriptstyle
2}}=\frac{mc}{\hbar}u.
$$
Then $\psi$ is such that
$$
\frac{\partial u}{\partial x^{\scriptscriptstyle 0}}=\frac{\partial
u}{\partial x^{\scriptscriptstyle 3}},\quad\frac{\partial u}{\partial
x^{\scriptscriptstyle 1}}=0.
$$
From this we have that
\begin{equation}\label{sp}
\psi=\left[\begin{array}{r} 1\\ 1\\ -1\\
1\end{array}\right]e^{\frac{mc}{\hbar}x^{\scriptscriptstyle
2}+f(x^{\scriptscriptstyle 0}+x^{\scriptscriptstyle
3})+ig(x^{\scriptscriptstyle 0}+x^{\scriptscriptstyle 3})}
\end{equation}
is a solution of the Dirac equation without gravitational field. Here $g(x^
{\scriptscriptstyle 0}+x^{\scriptscriptstyle 3})$ and $f(x^{\scriptscriptstyle 0}
+x^{\scriptscriptstyle 3})$ are smooth real functions.

\begin{figure} %
\centering
\includegraphics{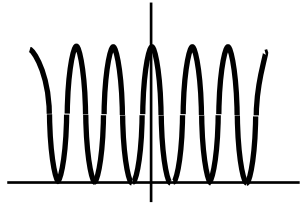}\\ 
\caption{The intensity of the distribution of
probability amplitude in point
$ 8e^{2\frac{mc}{\hbar}x^{\scriptscriptstyle 2}}=1,$ $x^{\scriptscriptstyle 0}
=0$.}
\label{pic0}
\end{figure}

From theorem~1 we get that (\ref{sp}) is a ghost spinor iff $g(x^{\scriptscriptstyle
0}+x^{\scriptscriptstyle 3})=const\in \R.$
We take the solutions for real wave
\begin{equation}\label{sp1}
\psi=\left[\begin{array}{r} 1\\ 1\\ -1\\
1\end{array}\right]e^{\frac{mc}{\hbar}x^{\scriptscriptstyle
2}+i(x^{\scriptscriptstyle 0}+x^{\scriptscriptstyle 3})}
\end{equation}
and for ghost wave
\begin{equation}\label{sp2}
\theta=\left[\begin{array}{r} 1\\ 1\\ -1\\
1\end{array}\right]e^{\frac{mc}{\hbar}x^{\scriptscriptstyle 2}}.
\end{equation}
As (\ref{sp1}) as (\ref{sp2}) have four-vector of the Dirac current
$$
j^{(k)}=(4e^{\frac{mc}{\hbar}x^{\scriptscriptstyle
2}},0,0,-4e^{\frac{mc}{\hbar}x^{\scriptscriptstyle 2}}).
$$
As both solutions
have alike current densities we can calculate the resulting wave after
collision of these particles. We shall find the square of modulus of probability
amplitude $\psi+\theta$. We have
\begin{equation}\label{intr}
{|\psi+\theta|}^2={(\psi+\theta)}^*(\psi+\theta)=8e^{2\frac{mc}{\hbar}x^{\scriptscriptstyle
2}}(1+\cos(x^{\scriptscriptstyle 0}+x^{\scriptscriptstyle 3})).
\end{equation}
If $x^{\scriptscriptstyle 0}$ and $x^{\scriptscriptstyle 2}$ are fixed we
get interference picture (Fig.\ref{pic0}). In result we have else one
variant of the interaction between ghost and real spinors.

\subsection{Two-slit experiment and ghost spinors}

\begin{figure} %
\centering
\includegraphics{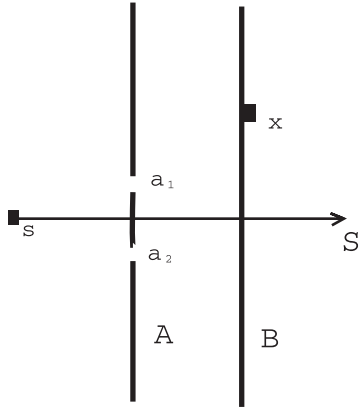}\\
\caption{Two-slit experiment.}
\label{pic4}
\end{figure}

In result we have that interaction between ghost and real spinors may be
differents. But we want to research namelly the participation of shadow
particles in two-slit experiment (Fig.\ref{pic4}).

Let the particle is in point $s$, on the screen $B$ in point $x$
particle detector is fixed. It defines an appearance of the particle at
the region of screen. On the screen $A$ the symmetricals about axis $S$
slits $a_1$ and $a_2$ are fixed. The wave spreads along axis $S$. The
distribution of the particles on the screen $B$ is interested. Let electrons
are emited on one. As known an interference pattern exist
as in this case as the flow is emited.

Let ${\psi}_1$ is probability amplitude of electron which passed
through slit $a_1$ and ${\psi}_2$ is probability amplitude of electron which
passed through slit $a_2$. In this case a distribution of the resulting
wave is defined by expression ${|{\psi}_1+{\psi}_2|}^2$. Let
interference exist, i.e. the distances between $a_1$ and $a_2$ and screens
$A$ and $B$ so that we have interleaving of maximums and minimums on
the screen $B$.

Let now ${\theta}_1$ is a shadow electron which passed through slit $a_1$
and ${\theta}_2$ is a shadow electron which passed through slit $a_2$.
Further we shall use the hypothesis that a shadow electron is a ghost
electron \cite{11,12}, i.e. a shadow electron is a solution of the Dirac
equation.

Now we shall use the khown indications of the quantum mechanics. Let the
state
$|y\rangle$ means that an electron in initial state in point $y$, the state $\langle
y|$ means that an electron in final state in point $y$. Herewith also let us
assume that symbol $\langle\:|\:\rangle$ means the resulting state of system
while experiment and also ${\langle \:|\:\rangle}_{\psi}$ and ${\langle
\:|\:\rangle}_{\theta}$ mean the states for real spinor field $\psi$
and ghost spinor $\theta$. In considering indications
\begin{equation}\label{p1}
{\psi}_1={\langle
x|{a}_1\rangle}_{{\psi}_1}{\langle{a}_1|s\rangle}_{{\psi}_1},
\end{equation}
\begin{equation}\label{p2}
{\psi}_2={\langle
x|{a}_2\rangle}_{{\psi}_2}{\langle{a}_2|s\rangle}_{{\psi}_2},
\end{equation}
\begin{equation}\label{tet1}
{\theta}_1={\langle
x|{a}_1\rangle}_{{\theta}_1}{\langle{a}_1|s\rangle}_{{\theta}_1},
\end{equation}
\begin{equation}\label{tet2}
{\theta}_2={\langle
x|{a}_2\rangle}_{{\theta}_2}{\langle{a}_2|s\rangle}_{{\theta}_2}.
\end{equation}
Let us define the probability distribution that a
real particle gets to point $x$ or through slit $a_1$ or through slit $a_2$
from starting point $s.$
This probability distribution is definded by the state
\begin{equation}\label{sost1}
{\langle x|s\rangle}_1={\langle x|
{a}_1\rangle}_{{\psi}_1}{\langle{a}_1|s\rangle}_{{\psi}_1}+{\langle
x|{a}_2\rangle}_{{\psi}_2}{\langle{a}_2|s\rangle}_{{\psi}_2}.
\end{equation}
Herewith we can not know which slit was passed by particle. We also not  take into
account the interaction between a real electron and ghost electrons.

The state ${\langle x|s\rangle}_2$ defines an intensity that a real electron
gets on the screen $B$ from  point $s.$ Herewith interactions between real
and shadow particles are taken into consideration and we can not know which slit
was passed by a real electron. Now we shall
calculate this state. There are four variant in experiment.
A real electron and a shadow electron can pass through one slit or different slits.
The states
${\langle{a}_1|s\rangle}_{{\theta}_2},\,{\langle{a}_2|s\rangle}_{{\theta}_1},\,$
 ${\langle{a}_1|s\rangle}_{{\psi}_2}$ and ${\langle{a}_2|s\rangle}_{{\psi}_1}$
are not possible. It follow from definition of wave functions
${\psi}_i$ and ${\theta}_i.$\footnote {In this case $i=1,2$.} Now we get
$$
{\langle
x|s\rangle}_2={\langle x|
{a}_1\rangle}_{{\psi}_1}{\langle{a}_1|s\rangle}_{{\psi}_1}{\langle x|
{a}_1\rangle}_{{\theta}_1}{\langle{a}_1|s\rangle}_{{\theta}_1}+{\langle
x|{a}_2\rangle}_{{\psi}_2}{\langle{a}_2|s\rangle}_{{\psi}_2}{\langle x|
{a}_2\rangle}_{{\theta}_2}{\langle{a}_2|s\rangle}_{{\theta}_2}+
$$
\begin{equation}\label{sost2}
+{\langle
x|{a}_1\rangle}_{{\psi}_1}{\langle{a}_1|s\rangle}_{{\psi}_1}{\langle x|
{a}_2\rangle}_{{\theta}_2}{\langle{a}_2|s\rangle}_{{\theta}_2}+{\langle
x|{a}_2\rangle}_{{\psi}_2}{\langle{a}_2|s\rangle}_{{\psi}_2}{\langle x|
{a}_1\rangle}_{{\theta}_1}{\langle{a}_1|s\rangle}_{{\theta}_1}.
\end{equation}
In result we use (\ref{p1}) -- (\ref{tet2}) and find
\begin{equation}\label{sost22}
{|{\langle x|s\rangle}_2|}^2={|{\psi}_1{\theta}_1+{\psi}_1
{\theta}_2+{\psi}_2{\theta}_1+{\psi}_2{\theta}_2|}^2={|{\psi}_1+{\psi}_2|}^2
\cdot {|{\theta}_1+{\theta}_2|}^2.
\end{equation}

Let us notice that we take into account the influence of shadow electrons to
the interference. If in this case we can not observe an interference pattern
then this influence is absent.
From this we conclude that we must show the existence of
interleaving of maximums and minimums at the function (\ref{sost22})
as a function which depend on point $x$ of the screen $B.$

\begin{figure} %
\centering
\includegraphics{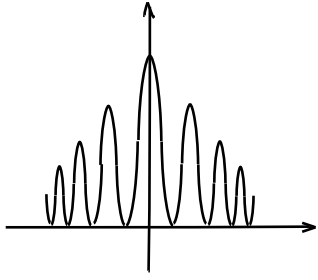}\\
\caption{The two-slit interference with dark centre.}
\label{pic1}
\end{figure}

As known the function ${|{\psi}_1+{\psi}_2|}^2$ is the square of modulus
of expression (\ref{sost1}) and it defines the intencity of appearance of an
electron to the screen $B.$
The interference pattern and interliving of maximums and minimums at
this function are simultaneously observed. For example the respective graph
hase the resemblance in kind with graph on (Fig.\ref{pic1}).
As known in an one-slit experiment the probability of appearance of an electron
to the screen $B$ is defined by Gauss distribution. So the wave functions
${\psi}_1$ and ${\psi}_2$ must be as \footnote
{In spite of the fact that we consider a case with (\ref{sost22}), we must
assume the existence of the interliving of maximums and minimums and for
function ${|{\psi}_1+{\psi}_2|}^2$. This function corresponds to variant
(\ref{sost1}) which not take into account the shadow particles. Since if the
hypothesis of the Deutsch shadow particles is right then, in approximation,
i.e. without shadow particles, the earlier existing explanation of the
quantum particle interference is must executed. And this means that expression
${|{\psi}_1+{\psi}_2|}^2$ hase the above-mentione type. So in case of shadow
particles existence the definition (\ref{p1p2})
of the amplitudes ${\psi}_1$ and ${\psi}_2$ is corrected.
}
\begin{equation}\label{p1p2}
{\psi}_1=u\cdot e^{-A{(x+d)}^2+i{\alpha}(x)},\quad
{\psi}_2=u\cdot e^{-A{(x-d)}^2+i{\beta}(x)}.
\end{equation}
Herewith $u^*u\neq 0$ and
$u$ is bispinor with complex numbers in components, $A\neq 0$ is not depended
by $x$, and ${\alpha}(x)$, ${\beta}(x)$ such that the square of modulus
of sum of the functions (\ref{p1p2}) is corresponded to observation of a
light-shadow pattern. Moreover $d$ is a half-distance between slits.

Let us also notice as shadow particles are particles in parallel
universes, much the same to our universe, then the wave ${\theta}_1$ must
have distribution such as distribution for ${\psi}_1$ and wave ${\theta}_2$
must have distribution such as distribution for ${\psi}_2$. Then from
results of theorem~1 we conclude that
\begin{equation}\label{t1t2}
{\theta}_1=u\cdot e^{-A{(x+d)}^2+i\cdot c_1},\quad
{\theta}_2=u\cdot e^{-A{(x-d)}^2+i\cdot c_2},\quad
c_i\in \R .
\end{equation}
It means that for sufficiently small $d$ the probability distribution
${|{\theta}_1+{\theta}_2|}^2$ is corresponded to absence of
interference pattern. This distribution is a Gauss distribution.

So we have the interliving of maximums and minimums
for graph of the function (\ref{sost22}). Moreover this function and
function ${|{\psi}_1+{\psi}_2|}^2$ have minimums in the same points,
but they have different maximum meanings. It is
right as function ${|{\theta}_1+{\theta}_2|}^2$ is not equal to zero, as
(\ref{t1t2}) is right. In result we have that interference pattern exists
if we take into account shadow electrons \cite{2} and
if we take that a
shadow electron same a ghost electron \cite{11,12}.

\section{Possibility of experimental test}

We considered a two-slit experiment (Fig.\ref{pic4}), the
existence of the Deutsch particles was taken into account. But in
corresponding experiment (Fig.\ref{pic4}) we assume that only one shadow
electron gets after screen $A$. But many shadow electrons can get after
screen $A$. What is an interference pattern in this case? Let us notice that
we can not define the number of shadow particles which get after the screen
$A$. Because we can not fixed shadow particles.

Let real electrons which passed through slits $a_1$ or $a_2$
are defined, as earlier, by wave functions ${\psi}_1$ or ${\psi}_2$.
The number of shadow electrons which passed through slits $a_1$ or
$a_2$ is equal to $n$. These shadow electrons are defined by wave functions
$\left\{{\theta}_1^{(m)}\right\}$ or $\left\{{\theta}_2^{(m)}\right\}$,
accordingly. Here $m=\overline{1,n}$. The wave functions have the type (\ref{p1p2}),
$\left\{{\theta}_1^{(m)}\right\}$ and $\left\{{\theta}_2^{(m)}\right\}$
are defined by (\ref{t1t2}), where $\forall\; m=\overline{1,n}\quad
c_i=c_i^m$. Then, as for formula (\ref{sost2}), we have the next expression
for square of modulus of probability amlitude
$$
{\langle x|s\rangle}_2=\left({\psi}_1{\theta}_1^{(1)}\cdot
...\cdot{\theta}_1^{(n)}+{\psi}_1{\theta}_1^{(1)}\cdot
...\cdot{\theta}_1^{(n-1)}{\theta}_2^{(n)}+{\psi}_1{\theta}_1^{(1)}\cdot ...
\cdot{\theta}_1^{(n-2)}{\theta}_2^{(n-1)}{\theta}_1^{(n)}+\right.
$$
\begin{equation}\label{e1}
\left.+{\psi}_1{\theta}_1^{(1)}\cdot
...\cdot{\theta}_1^{(n-2)}{\theta}_2^{(n-1)}{\theta}_2^{(n)}+...+{\psi}_2{\theta}_1^{(1)}\cdot
...\cdot{\theta}_1^{(n)}+...\right).
\end{equation}
Now we convert (\ref{e1}) and get
$$
{\langle x|s\rangle}_2=\left({\psi}_1{\theta}_1^{(1)}\cdot
...\cdot{\theta}_1^{(n-1)}+{\psi}_1{\theta}_1^{(1)}\cdot
...\cdot{\theta}_2^{(n-1)}+...+{\psi}_2{\theta}_1^{(1)}\cdot
...\cdot{\theta}_1^{(n-1)}+\right.
$$
$$
+\left. {\mathstrut}^{\mathstrut}...\right){\theta}_1^{(n)}
+\left({\psi}_1{\theta}_1^{(1)}\cdot
...\cdot{\theta}_1^{(n-1)}+{\psi}_1{\theta}_1^{(1)}\cdot
...\cdot{\theta}_2^{(n-1)}+...\right.
$$
$$
...\left. +{\psi}_2{\theta}_1^{(1)}\cdot
...\cdot{\theta}_1^{(n-1)}+...\right){\theta}_2^{(n)}.
$$
From this we have
$$
{\langle x|s\rangle}_2=\left({\psi}_1{\theta}_1^{(1)}\cdot
...\cdot{\theta}_1^{(n-1)}+{\psi}_1{\theta}_1^{(1)}\cdot
...\cdot{\theta}_2^{(n-1)}+...+{\psi}_2{\theta}_1^{(1)}\cdot
...\cdot{\theta}_1^{(n-1)}+...\right.
$$
\begin{equation}\label{e2}
\left. ...+{\psi}_2{\theta}_2^{(1)}\cdot
...\cdot{\theta}_2^{(n-1)}\right)\left({\theta}_1^{(n)}+{\theta}_2^{(n)}\right).
\end{equation}
If we shall consecutively run this prosedure for expression (\ref{e2})
we get
\begin{equation}\label{e3}
{|{\langle x|s\rangle}_2|}^2={|{\psi}_1+{\psi}_2|}^2
\cdot \prod\limits_{i=1}^n{|{\theta}_1^{(i)}+{\theta}_2^{(i)}|}^2,
\end{equation}
here $n$ is the number of shadow particles which passed through slits
after screen $A$.
The intensity of appearance of electrons on the screen $B$ is defined by
expression (\ref{e3}).

We shall take into account the types of the functions ${\theta}_1^{(i)}$
and ${\theta}_2^{(i)}$. For every number $i$ functions
$|{\theta}_1^{(i)}+{\theta}_2^{(i)}|$ are defined by Gauss distribution if
a distance between slits is sufficiently small.
As earlier we conclude that expression
(\ref{e3}) defines some interference pattern. Also as earlier we notice that
the minimums of corresponding functions are in same points too.

Let us see to the getting results with standpoint of the experimental data
existence. In all experiments of quantum particles interference
we have
that the getting with experiment intensity of an electrons distribution
corresponds to the function
${|{\psi}_1+{\psi}_2|}^2$ which is multiplied on the scale factor. It is right
as ${|{\psi}_1+{\psi}_2|}^2$ must be a function of distribution.

So A.K. Guts suggest to  involve the multiplying on the
product
\begin{equation}\label{prod}
\prod\limits_{i=1}^n{|{\theta}_1^{(i)}+{\theta}_2^{(i)}|}^2
\end{equation}
with multiplying on a scale factor. Every new electron
will adds a new positive factor. If in (\ref{prod}) the coinverses
factors meet then their influence is canceled. So if we consider infinite
number of shadow particles, i.e. we let $n=\infty,$ the  multiplier
(\ref{prod}) will be equal to number one. We have that intensity of
particles distribution with consideration shadow particles is equal to
intensity of distribution of particles without one.

As known slits have a finite size and electrons have a finite size too. So
on the experiment result only finite number of ghost particles are acted, i.e.
$n\neq\infty.$ It is we observe in experiment: scale factors in
different cases are differed.

Can we test the existence of shadow particles from received data? This
probability exists. For this the checking of shadow electrons which influence
on experiment is sufficient condition. The problem is that we can not fixed
the Deutsch particles with some particle detector. But may be a solution of
this problem in next.

\begin{figure} %
\centering
\includegraphics{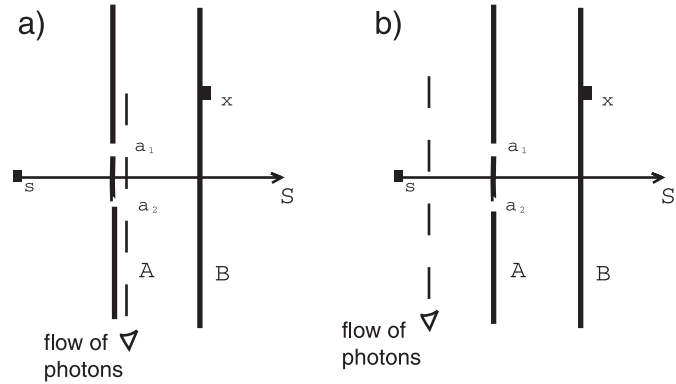}\\
\caption{ Here a) classical experiment, b) proposed experiment.}
\label{pic5}
\end{figure}

Actually the knowlege about number of shadow particles which passed through slits
$a_1$ and $a_2$ is not necessities condition. Let us remember to one of
the which way experiment (Fig.\ref{pic5}a). For this between screens $A$ and
$B,$ as closer to screen $A$ as possible, the flow of photons is emited. As
known the information about system is conteined in this measurement and in
this case we have a disappearance of interference picture. It is the
measurement act on the result of experiment. Let us see on this case with view
of existence of Deutsch particles. When this measurement is conducted the
interaction between system and environment is conducted too. But interaction
between system and environment is interaction between Deutsch particles
and measurement which we take. It is explained by further. If the flow of
photons is not emited the interference picture exist as shadow and real
particles are fased \cite{2}. But if we have a which way experiment, i.e.
the flow of photons is measurement which act on observed result, the
shadow electron is repulsed by some shadow photon and the conflict between
these particles can not proceed. This discourse is correst as the formal
model of Multiverse exist and shadow photons was found \cite{3,4}
(these problems was not solved in \cite{2,11,12}). Moreover we assume that
may be than closer flow of photons to screen $A$ that less of shadow
electrons interact with real electron and, as result, less multipliers in
factor (\ref{prod}). From this imaginary experiment we conclude that may
be we can control, with some way, a number of shadow particles which act
to interference pattern.

Let now before screen $A$ the flow of photons is emited (Fig.\ref{pic5}b).
We assume that the number of shadow electrons which act on the result of
experiment is redused by some movement of flow of photons to the right along
axis $S$ and it is enlarged by some movement of flow of photons to the left
along one. Let we fixed some point of intersection of flow of photons with
axis $S$ and conducted the experiment. In result we have some interference
picture. Then we little move the source of this photons flow along axis $S$
to the right, for example, and take some interference pattern too. Then the
differention between values of two nearby maximums in second case must be
less then differention between values of two nearby maximums on first
interference pattern. It is follow from our results and assumption that flow
of photons control the number of Deutsch particles which act on
interference, in above described meaning.
If in offered experiment this result will fixed we can speak about existence
of shadow particles and about existence of Guts-Deutsch Multiverse.
We notice when flow of photons will be
sufficiently near to screen $A$ for that the information about way of
particle will be received the influence of shadow particles is greatly
reduced and interference pattern will be disappeared. Herewith as we shall
receive the information about way of particle the probability amplitude
will be satisfy no expression (\ref{e3}) but next expression:
\begin{equation}\label{e4}
{|{\langle x|s\rangle}_2|}^2=\left({|{\psi}_1|}^2+{|{\psi}_2|}^2\right)
\cdot \prod\limits_{i=1}^n{|{\theta}_1^{(i)}+{\theta}_2^{(i)}|}^2.
\end{equation}

So this sequence of experiments is that when source of flow of photons move
along axis $S$ (Fig.\ref{pic5}b)  the interference pattern is probed.

Except experiment (Fig.\ref{pic5}b) we can see else one experiment which was
suggested by M.S. Shapovalova. In this case the number of
shadow particles which act on an interference pattern is changed by sizes of
slits: than smaller the slits that smaller the number of these particles.
But in this case the function ${|{\psi}_1+{\psi}_2|}^2$ also will be
changed. But we think that we can receive some results about interaction
shadow and real particles and in this case.

\section*{Conclusion}

In this paper the some questions of development of Theory of Multiverse are
considered. We assumed that our reality is the Guts-Deutsch Multiverse \cite{2,3,4}.
In this aspect the interference between quantum particles is described. As know
in this time the problem of quantum mearsurements is researched. The
Deutsch shadow particles contribute some essential explanations to this
problem. In first, the some aspects of quantum particles interference are
explained by Deutsch \cite{2}. In this article we got that shadow electrons
can really act to the interference and this may be tested by some experiment, i.e.
we considered the experiment which may to solve the problem which theory is
more exact: Theory of Universe or Theory of Multiverse. One is contained in
Theory of Multiverse. But can we describe our world by this theory? Only
the experiment will defines this.


\end{document}